\documentclass{appolb}
\usepackage{epsfig}
\usepackage{showlabels}
\usepackage{amsmath}
\usepackage{amsfonts}
\usepackage{amssymb}

\newcommand{\beq}{\begin{equation}}
\newcommand{\eeq}{\end{equation}}
\newcommand{\ba}{\begin{array}{ccc}}
\newcommand{\ea}{\end{array}}

\newcommand{\be}{\begin{equation}}
\newcommand{\ee}{\end{equation}}
\newcommand{\bea}{\begin{eqnarray}}
\newcommand{\eea}{\end{eqnarray}}
\newcommand{\nn}{\nonumber}

\newcommand{\one}{\mbox{\bf 1}}

\def\Tr{{\mbox{Tr}}}

\def\hchi{\hat{\chi}}
\def\hH{\hat{H}}

\begin{document}
% \eqsec  % uncomment this line to get equations numbered by (sec.num)
\title{
Chiral Random Two-Matrix Theory and QCD with imaginary chemical potential
\thanks{Presented at the ESF Exploratory Workshop on ``Random Matrix Theory:
  From Fundamental Physics to Applications'' in
Krakow 
May 2007, Poland}%
% you can use '\\' to break lines
}
\author{G. Akemann
\address{Department of Mathematical Sciences \& BURSt Research Centre\\
Brunel University West London,
Uxbridge UB8 3PH, United Kingdom}
%\and
%the Name(s) of other Author(s)
%\address{and their affiliation}
}
\maketitle
\begin{abstract}
We summarise recent results for the chiral Random Two-Matrix Theory constructed
to describe QCD in the epsilon-regime with imaginary chemical
potential. The virtue of this theory is that unquenched Lattice simulations
can be used to determine both low energy constants $\Sigma$ and $F$ 
in the leading order chiral Lagrangian, due to their respective coupling to
quark mass and chemical potential. 
We briefly recall the analytic formulas for all density and individual
eigenvalue 
correlations and then illustrate them in detail in the simplest,
quenched  case with imaginary isospin chemical potential. 
Some peculiarities are pointed out for this example: 
i) the
factorisation of density and individual eigenvalue correlation
functions for large chemical potential
and ii) the 
factorisation of the non-Gaussian weight function of bi-orthogonal
polynomials into Gaussian weights with ordinary orthogonal polynomials.

\end{abstract}

\section{Introduction}
Non-Hermitian Random Matrix Theory (RMT) has received a lot of interest in the
past few years due to its relation to QCD with chemical potential $\mu$, 
see \cite{Amu07} a for recent review (and \cite{Jackra} this workshop). 
Many results have been obtained, including correlation functions 
\cite{KimJac03,James,AOSV},
individual eigenvalues \cite{ABSW} 
or the phase of the Dirac operator \cite{KimJac07},
and have been successfully compared to QCD lattice data \cite{AW,ABSW,KimBen}. 
We have now understood that RMT with (or without) chemical potential is
equivalent to QCD \cite{BA}
in the limit of the epsilon-regime of chiral Perturbation
Theory (echPT) \cite{GL}. 

The virtue of having $\mu\neq0$ is that it couples to $F$ to leading order in
echPT \cite{DomJac}. 
The downside of 
$\mu\neq0$ is of course the sign problem, making unquenched simulations very
hard. 
It was therefore 
proposed in \cite{DHSST} to use imaginary $\mu$ instead to determine 
$F$, keeping the Dirac operator eigenvalues real and thus making unquenched
simulations possible. First results for the two-point density 
\cite{DHSST} were derived directly from 
echPT for imaginary isopin chemical potential and compared to quenched and
unquenched Lattice data (of course real isopin chemical potential could also
be simulated unquenched). 
This inspired us to write down and solve the corresponding two-Matrix Theory
(2RMT) \cite{ADOS}, 
where in addition 
partial quenching is possible by setting one of the two $\mu_j$ to zero. 
This method has already been successfully compared to the lattice QCD in
\cite{DHSST,DeG}.  
In \cite{ADOS}, all unquenched density correlation functions were computed 
(including those for the non-chiral theory for QCD in three dimensions). This
lead to the construction of individual eigenvalues as well \cite{ADlat07}, and
has subsequently been proven to be equivalent to the corresponding echPT 
for all correlation functions \cite{BA}. 

The purpose of this paper is to illustrate the mathematical structure of these
results by using the simplest possible setting, the quenched theory with 
imaginary $\mu$ of isopin type. For results in full
generality, including partially quenched and unquenched examples we refer to 
\cite{ADOS,ADlat07}. 

The quenched case furthermore helps to point out the differences 
between correlations of two anti-Hermitian Dirac operators with 
imaginary $\mu$ isospin and eigenvalues on 
$\mathbb{R}^2$, and one non-Hermitian Dirac operator with 
real $\mu$ and eigenvalues on 
$\mathbb{C}$. 
Below we show that 
in the limit of large imaginary $\mu$ all correlation functions 
{\it factorise and become $\mu$-independent}, given by 
the product of two single, uncoupled Dirac operators. 
For large real $\mu$ however, the complex eigenvalue densities stays
{\it $\mu$-dependent} and becomes {\it rotationally invariant} 
around the origin in $\mathbb{C}$ \cite{AOSV}. 

This article is organised as follows. In the next section \ref{rmt} we recall
the 
2RMT and its equivalent echPT, as well as the general results for all
correlation functions. In section \ref{quench} we then specify 
these results to the simplest, quenched example with imaginary 
isospin, including
two interesting properties. First, the factorisation of all correlation
functions  
for large $\mu$ is derived and illustrated with 
several figures.
Second, the factorisation of the non-Gaussian 
weight on $\mathbb{R}^2$ into two Gaussian weights on $\mathbb{R}$ is shown.

%%%%%%%%%%%%%%%%%%%%%%%%%%%%%%%%%%%%%%%%%%%%%%%%%%%%%%%%%%%%%%%%%%%%%%%%

\section{RMT and echPT}\label{rmt} 

We begin by writing down the partition function of echPT given by \cite{DomJac}
\be
{\cal Z}
={\int_{U(N_f)} dU \det[U]^\nu}
\exp\left[{\Tr \frac{1}{4}{F^2V}{[U,B][U^\dagger, B]}
+ \Tr\frac12V \Sigma{M}( {U}+ {U^\dagger})}\right]\ .
\label{ZechPT}
\ee
Here $F$ and $\Sigma$ are the Pion decay constant and chiral condensate,
respectively. In the epsilon regime they have
as source terms chemical potential through the charge matrix 
$B=$diag$(\mu_1\one_{N_1},\mu_2\one_{N_2})$, and the diagonal mass matrix 
$M=$diag$(\{m_{f1}\},\{m_{f2}\})$, respectively.

For this theory all eigenvalue correlation functions are known
\cite{ADOS,ADlat07} and are equivalent \cite{BA} 
to the chiral 2RMT 
eq. (\ref{ZNf}). Deriving correlation functions from echPT one has to add
auxiliary fermion-boson pairs to generate the corresponding resolvents, and we
refer to \cite{BA} for details. In fact the RMT-echPT equivalence holds for
any number of chemical potentials, but only for {\it two} different chemical
potentials this theory has
been solved. From now on we set $\mu_1=-\mu_2$ for simplicity and
follow the 2RMT framework as it is much simpler. 
The corresponding partition function is defined as 
\be
{\cal Z}_{RMT}=\!\int d\Phi  d\Psi~ e^{-{N}{\rm Tr}\left(\Phi^{\dagger}
\Phi + \Psi^{\dagger}\Psi\right)}
\prod_{f1=1}^{N_1} \det[{\cal D}_++ m_{f1}] 
\prod_{f2=2}^{N_2} \det[{\cal D}_-+ m_{f2}] .
\label{ZNf}
\ee
The two anti-hermitian Dirac matrices ${\cal D}_\pm$ 
are given in terms of two complex, rectangular 
random matrices $\Phi$ and $\Psi$ of size $N\times (N+\nu)$
\bea
{\mathcal D}_\pm = \left( \begin{array}{cc}
0 & i \Phi \pm i \mu \Psi \\
i \Phi^{\dagger} \pm i \mu \Psi^{\dagger} & 0
\end{array} \right) ~.
\eea
When rotating to the eigenvalues $x_j$ and $y_j$ 
of  ${\cal D}_\pm$ the two random matrices
get coupled, leading to a non-trivial dependence on the unitary rotations. 
Integrating them out 
we obtain the following non-Gaussian eigenvalue model \cite{ADOS}, 
up to an overall constant, 
\bea
{\cal Z}_{RMT}
&=& \prod_{i=1}^N\left(\int_0^{\infty} dx_idy_i (x_iy_i)^{\nu+1}
\prod_{f1=1}^{N_1} (x_i^2+m_{f1}^2)
\prod_{f2=1}^{N_2} (y_i^2+m_{f2}^2) \right) 
\label{evrep}\\
\times&&\!\!\Delta_N(\{x^2\})\Delta_N(\{y^2\})
\det_{j,k}\left[I_{\nu}\left(\frac{1-\mu^2}{2\mu^2}N x_j y_k\right)\!
\right] 
e^{-\frac{N}{4\mu^2}(1+\mu^2) \sum_i x_i^2 +y_i^2 }
\nn
\eea
For later convenience we 
abbreviate the integrand or joint probability distribution function by 
${\cal P}(\{x\},\{y\})$.

%%%%%%%%%%%%%%%%%%%%%%%%%%%%%%%%%%%%%%%%%%%%%%%%%%%%%%%%%%%%%%%%%%%%%%%%%%%
\subsection{Definitions and Results}

If we define the weight function 
\bea
w(x,y)&\equiv& 
(xy)^{\nu+1}\prod_{f1=1}^{N_1}(x^2+m_{f1}^2)
\prod_{f2=1}^{N_2}(y^2+{m}_{f2}^2)\ 
\nn\\
&\times&I_{\nu}\left(\frac{(1-\mu^2)}{2\mu^2}Nxy\right) 
e^{-\,\frac{N}{4\mu^2}(1+\mu^2)(x^2+y^2)}
\label{weight}
\eea
we can find the corresponding bi-orthogonal polynomials
\beq
\label{biop}
\int_0^{\infty} dx dy \ w(x,y)\ 
P_n(x^2)\ 
Q_k(y^2)
= h_n\delta_{nk} ~, 
\eeq
that depend parametrically on the masses. All correlation functions 
defined in eqs. (\ref{allRnk}), (\ref{Ekl})  and (\ref{pkl}) 
below can then be expressed in terms of the 4 kernels $K_N,\ H_N,\
\hat{H}_N$ and $M_N$ that are constructed respectively
from the two bi-orthogonal polynomials $P_k$ and $Q_k$,
the polynomials $P_k$ and
the generalised Bessel transform of its partner
\be
\hchi_k(x) \equiv \int_0^{\infty}dy\ w(x,y) Q_k(y^2)\ , 
\label{chidef}
\ee
the polynomial $Q_k$ and its partners transform, and both transforms.
The density correlation functions are then given by \cite{ADOS}
\bea
&&R_{k,l}(x_{1,\ldots,k},\ y_{1,\ldots,l})\equiv
\frac{N!^2}{{\cal Z}(N-k)!(N-l)!}
\int_0^{\infty}\!\!  \prod_{i=k+1}^N \!\!dx_i \!\prod_{j=l+1}^N \!\!dy_j 
{\cal P}(\{x\},\{y\})
\nn\\
&&=\det_{1\leq i_1,i_2\leq n;\ 1\leq  j_1,j_2\leq k}
\left[
\begin{array}{cc}
H_N(x_{i_1},x_{i_2}) &
M_N(x_{i_1},y_{j_2})-w(x_{i_1},y_{j_2})\\ 
K_N(y_{j_1},x_{i_2}) & \hH_N(y_{j_1},y_{j_2})\\
\end{array}
\right].
\label{allRnk}
\eea
The simplest nontrivial example is the density $R_{1,1}(x,y)$
to find an eigenvalue of
${\cal D}_+$  at $x$ and of ${\cal D}_-$  at $y$.
When all eigenvalues of one kind are integrated out one finds
back the densities of the one-Matrix Theory (1RMT), 
which are then $\mu$-independent.

Alternatively to the density correlations one can define the so-called 
gap probability that
the interval 
$[0,s]$ is occupied by $k$ eigenvalues 
and $[s,\infty)$ by $(N-k)$ eigenvalues of ${\cal D}_+$, 
and that the interval 
$[0,t]$ is occupied by $l$ eigenvalues 
and $[t,\infty)$ by $(N-l)$ eigenvalues of ${\cal D}_-$:
\bea
E_{k,l}(s,t) &\equiv& \frac{N!^2}{{\cal Z}(N-k)!(N-l!)}
\int_0^s dx_1\ldots dx_{k}\int_s^\infty dx_{k+1}\ldots dx_N 
\label{Ekl}\\
&&\times
\int_0^t dx_1\ldots dx_{l}\int_t^\infty dy_{l+1}\ldots dy_N 
{\cal P}(\{x\},\{y\})
\nn\\
&=& \sum_{i=0}^{N-k}\sum_{j=0}^{N-l} \frac{(-)^{i+j}}{i!j!} 
\int_0^s  dx_1\ldots dx_{k+i}\int_0^t  dy_1\ldots dy_{l+j}
R_{k+i,l+j}\ . \ \ \ \ \label{EklR}
\eea
Here we have also given its expansion in terms of density correlations
\cite{ADlat07}. Obviously if all densities are known all gap probabilities
follow, and vice versa. Taking the following derivatives of the gap
probabilities 
\be
 \frac{\partial^2E_{k,l}(s,t)}{\partial s\partial_t} 
= k!\ l!\left( p_{k,l}(s,t) - p_{k+1,l}(s,t)
-p_{k+1,l}(s,t) + p_{k+1,l+1}(s,t)\right),
\label{Eklpkl}
\ee
then leads to individual eigenvalue distributions defined as
 \bea
p_{k,l}(s,t) &\equiv& \frac{kl}{{\cal Z}}
 {N \choose k} {N \choose l}
\int_0^s dx_1\ldots dx_{k-1}\int_s^\infty dx_{k+1}\ldots dx_N 
\label{pkl}\\
&&\times
\int_0^t dy_1\ldots dy_{l-1}\int_t^\infty dy_{l+1}\ldots dy_N 
{\cal P}(x_k=s,y_l=t)\ .
\nn
\eea
Following eq. (\ref{EklR}) they can be expanded in terms of densities as well,
and we will use this expansion below (see eq. (\ref{p00ex})).

%%%%%%%%%%%%%%%%%%%%%%%%%%%%%%%%%%%%%%%%%%%%%%%%%%%%%%%%%%%%%%%%%%%%%%%%%%%

\section{The Quenched Theory: Illustrations and Peculiarities}\label{quench}

In the following we illustrate the above results with the simplest quenched 
example, taking $N_1=N_2=0$. From symmetry the bi-orthogonal polynomials 
become equal, $P_k=Q_k$, and they 
are simply given by
Laguerre polynomials. Their Bessel transforms become the wave functions of
the Laguerre polynomials, where details are given in the next subsection 
\ref{wfac}. In particular two of the kernels then coincide, $H_N=\hat{H}_N$.

The large-$N$ limit can easily be taken using the standard Bessel asymptotic
of Laguerre polynomials.  We keep 
\bea
\alpha^2&\equiv&\lim_{N\to\infty}2N\mu^2 \ (\  =\ VF^2\mu_1^2)\nn\\
\xi_k &\equiv& \lim_{N\to\infty}Nx_k\ (\  =\ V\Sigma x_k) 
\ \ ,\ 
\zeta_k \ \equiv\ \lim_{N\to\infty}Ny_k\ (\  =\ V\Sigma y_k)\ , 
\label{limN}
\eea
fixed where in parenthesis the corresponding echPT quantities are given. 
Masses are rescaled as the eigenvalues when present. 
The limit eq. (\ref{limN}) results into the following building blocks for the
correlation functions, the microscopic kernels:
\be
{\cal I}^\kappa(\xi,\zeta)\equiv \int_0^1dt\,t\ e^{\,\kappa\frac12\alpha^2t^2}
J_\nu(\xi t)J_\nu(\zeta t)\ ,
\label{kernel}
\ee
where $\kappa=+1,0,-1$ for the limit of $K_N$, $H_N$ and $M_N$ respectively. 
The simplest non-trivial example is the rescaled density $R_{1,1}(x,y)$:
\be
\rho_{1,1}(\xi,\zeta)=
{\cal I}^0(\xi,\xi){\cal I}^0(\zeta,\zeta)-
\xi\zeta{\cal I}^+(\xi,\zeta)
\left
({\cal
  I}^-(\xi,\zeta)-\frac{I_\nu\left(\frac{\xi\zeta}{\alpha^2}\right) }{\alpha^2}
e^{-\frac{\xi^2+\zeta^2}{2\alpha^2}}
\right).
\label{r11}
\ee
\begin{figure}[-h]
\centerline{\epsfig{figure=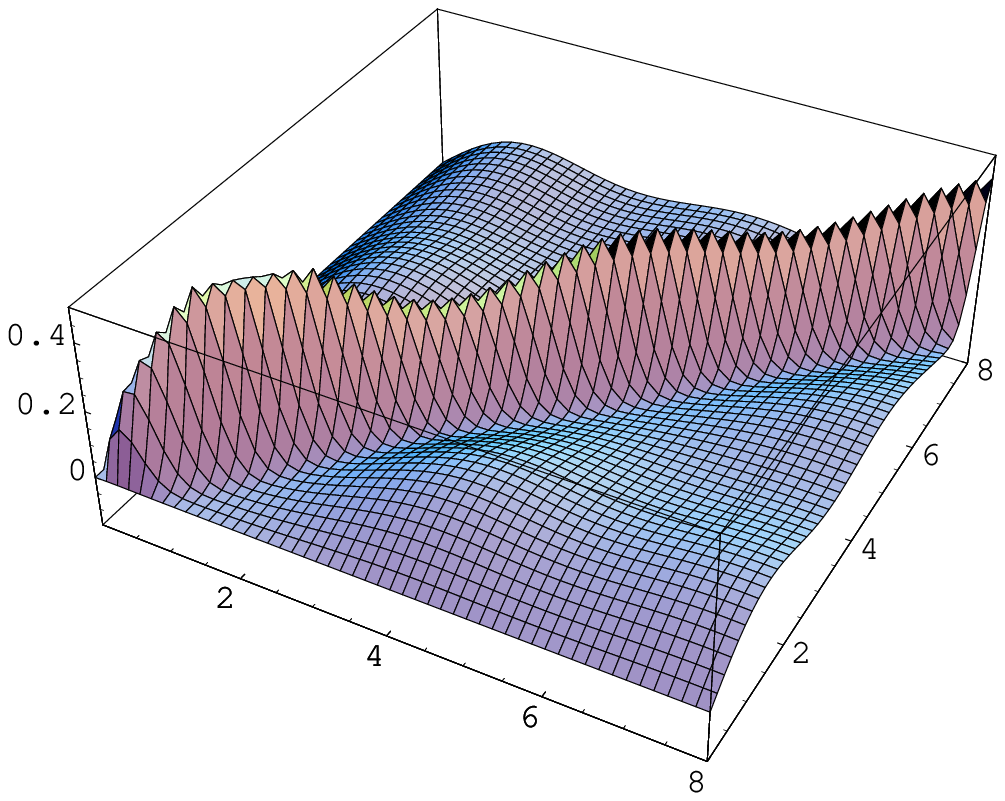,width=15pc}
\epsfig{figure=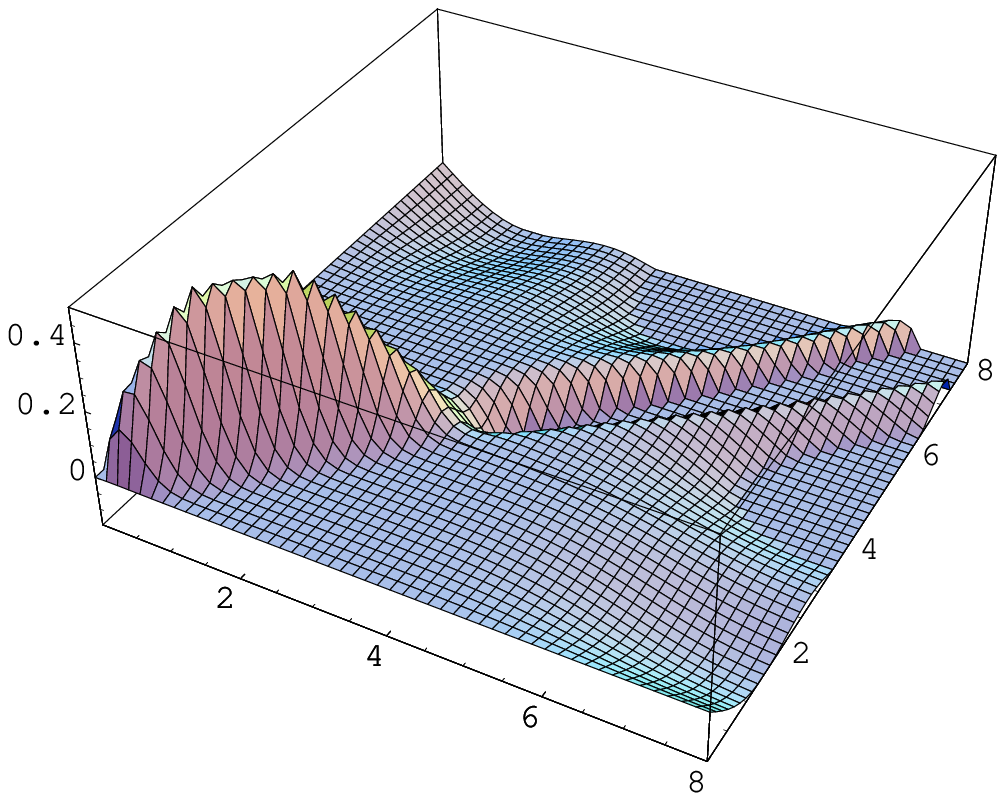,width=15pc}
\put(-360,140){$\rho_{1,1}(s,t)$}
\put(-180,140){$p_{1,1}(s,t)$ eq.(16)}
}
\centerline{\epsfig{figure=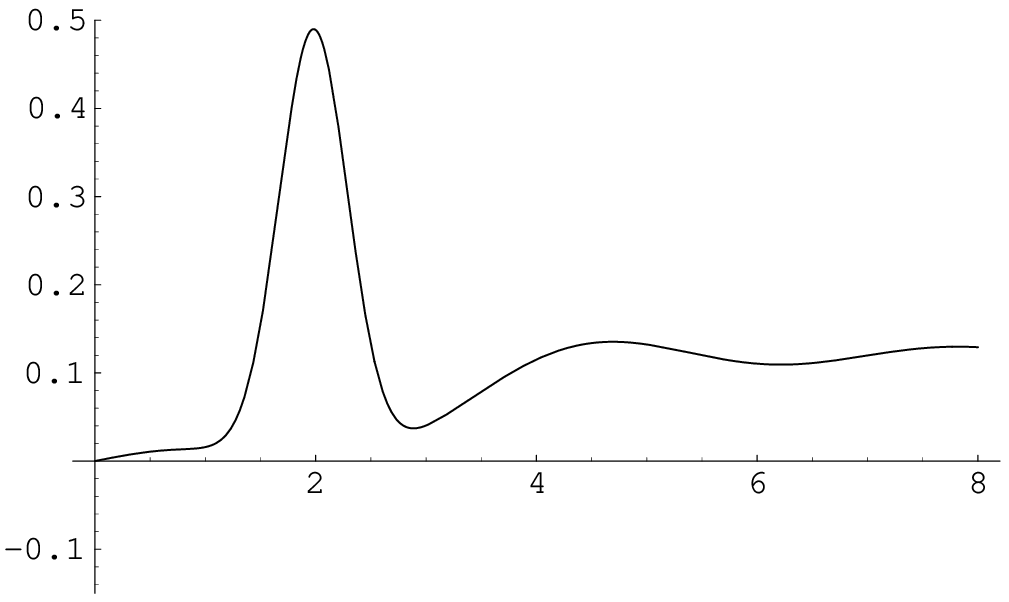,width=15pc}
\epsfig{figure=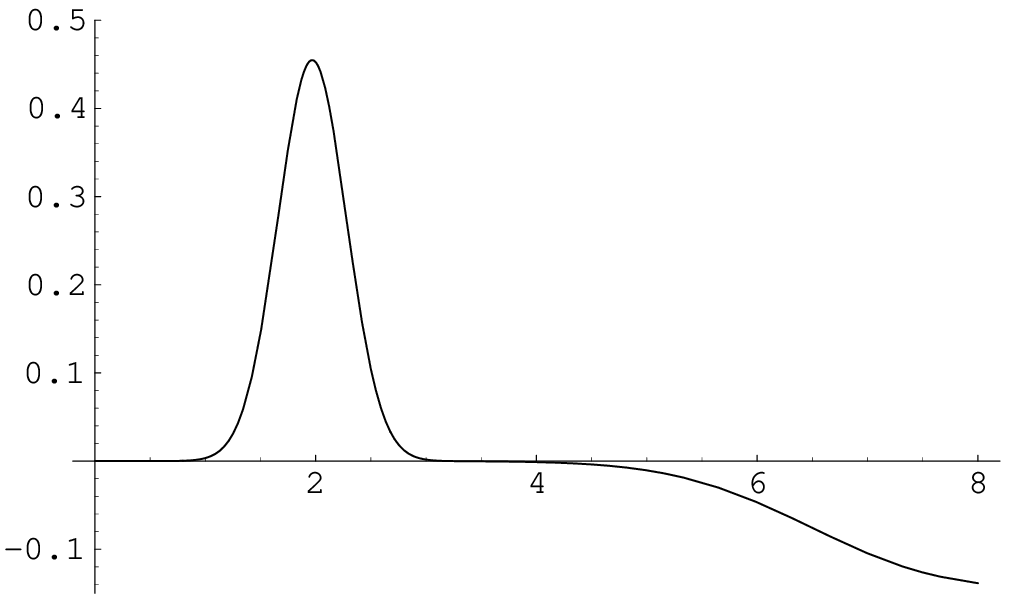,width=15pc}
\put(-280,80){$\rho_{1,1}(s=2,t)$}
\put(-250,0){$t$}
\put(-100,80){$p_{1,1}(s=2,t)$ eq. (16)}
\put(-70,0){$t$}}
\caption{The quenched
density $\rho_{1,1}(s,t)$ 
(top left) vs the individual eigenvalue distribution
$p_{1,1}(s,t)$ (top right)
expanded to the order given in eq. (\protect{\ref{p00ex}}), both 
at $\nu=0$ and $\alpha=0.159$. 
The fact that $p_{1,1}(s,t)$ becomes negative is an artefact of the
approximation. Higher order terms will ensure that it remains zero at large
distance from the origin.
The lower plots show corresponding 2D cuts at fixed $s=2$. The advantage of
$p_{1,1}(s,t)$ is that it is localised without background.
}
\label{R11quenchnu0}
\end{figure} 
For the corresponding individual eigenvalue distribution we use the expansion
following from eq. (\ref{EklR})
\be
p_{1,1}(s,t)\ =\ \rho_{1,1}(s,t)\ -\  \int_0^s dx\, \rho_{2,1}(x,s,t)
\ -\ \int_0^t dy\, \rho_{1,2}(s,t,y)\ +\ \ldots\ .
\label{p00ex}
\ee
Both eqs. (\ref{r11}) and (\ref{p00ex}) are displayed in
fig. \ref{R11quenchnu0}. 
For comparison we display the same quantities of the $\mu$-independent 1RMT. 
Its rescaled density
reads
\be
\rho_{1}(\xi)
=\frac{\xi}{2}\left[J_{\nu}^2(\xi)
- J_{\nu+1}(\xi)J_{\nu-1}(\xi)\right] 
\ (\ =\ {\cal I}^0(\xi,\xi)= \rho_{1,0}(\xi)=\rho_{0,1}(\xi)\ )\ ,\ 
\label{rhoQ}
\ee
\begin{figure}[-h]
\includegraphics[width=.5\textwidth]{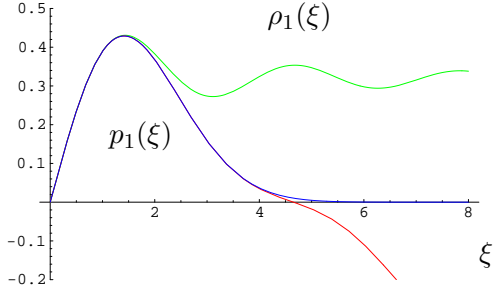}
\put(-80,100){$\rho_{1}(\xi)$}
\put(-140,55){$p_{1}(\xi)$}
\put(0,10){$\xi$}
\caption{1RMT at $\nu=0$:  
the quenched density 
eq. (\ref{rhoQ}) (green),
the exact distribution of the first eigenvalue 
eq.  (\protect{\ref{p1}}) (blue),  and the expansion  \cite{ADp}
corresponding to eq.  (\protect{\ref{p00ex}}) (red).
}
\label{rhoreal}
\end{figure} 
see fig. \ref{rhoreal}. There we include both the exact distribution of first
eigenvalue 
\cite{DNW}
\be
p_1(\xi)=\frac12\xi\ e^{-\frac14 \xi^2} \ \ (\nu=0)\ , 
\label{p1}
\ee
and its corresponding approximation \cite{ADp}.

%%%%%%%%%%%%%%%%%%%%%%%%%%%%%%%%%%%%%%%%%%%%%%%%%%%%%%%%%%%%%%%%%%%%%%%%%%%

\subsection{Factorisation of correlation functions}

In the limit of large chemical potential, $\alpha\gg 1$, the quenched density
correlation functions factorise,
\be
\lim_{\alpha\gg1}\rho_{n,k}(\xi_{1,\ldots,n},\zeta_{1,\ldots,k}) =
 \prod_{i}^n\xi_i \rho_{n,0}(\xi_{1,\ldots,n})\ \prod_j^k \zeta_j
\rho_{0,k}(\zeta_{1,\ldots,k})\ , 
\label{factorR}
\ee
where the two factors are given by the $\mu$-independent 1RMT
quantities 
\be
\rho_{n,0}(\xi_{1,\ldots,n})=\rho_{0,n}(\xi_{1,\ldots,n})=
\det_{i,j}[{\cal I}^0(\xi_i,\xi_j)] \ .
\label{rhok}
\ee
This follows from the vanishing of the upper right corner in the determinant 
eq. (\ref{allRnk}) when $\alpha$ is large: the corresponding 
microscopic kernel $M_N$ converges to
the weight in this limit 
\be
\lim_{\alpha\gg1}
\left({\cal
  I}^-(\xi,\zeta)-\frac{I_\nu\left(\frac{\xi\zeta}{\alpha^2}\right) }{\alpha^2}
e^{-\frac{\xi^2+\zeta^2}{2\alpha^2}}\right)=0\ ,
\label{Mtow}
\ee
keeping $\xi/\sqrt{\alpha}$ finite. 
Thus the density from our example eq. (\ref{r11}) factorises, as is shown in
fig. \ref{R11factor} left. 
The deeper reason for the limit eq. (\ref{Mtow}) will become
clearer in the next subsection \ref{wfac}.

The factorisation of the densities leads to factorised gap
probabilities and individual eigenvalue distributions as well, as follows
from eq. (\ref{EklR}): 
\begin{figure}[-h]
\centerline{\epsfig{figure=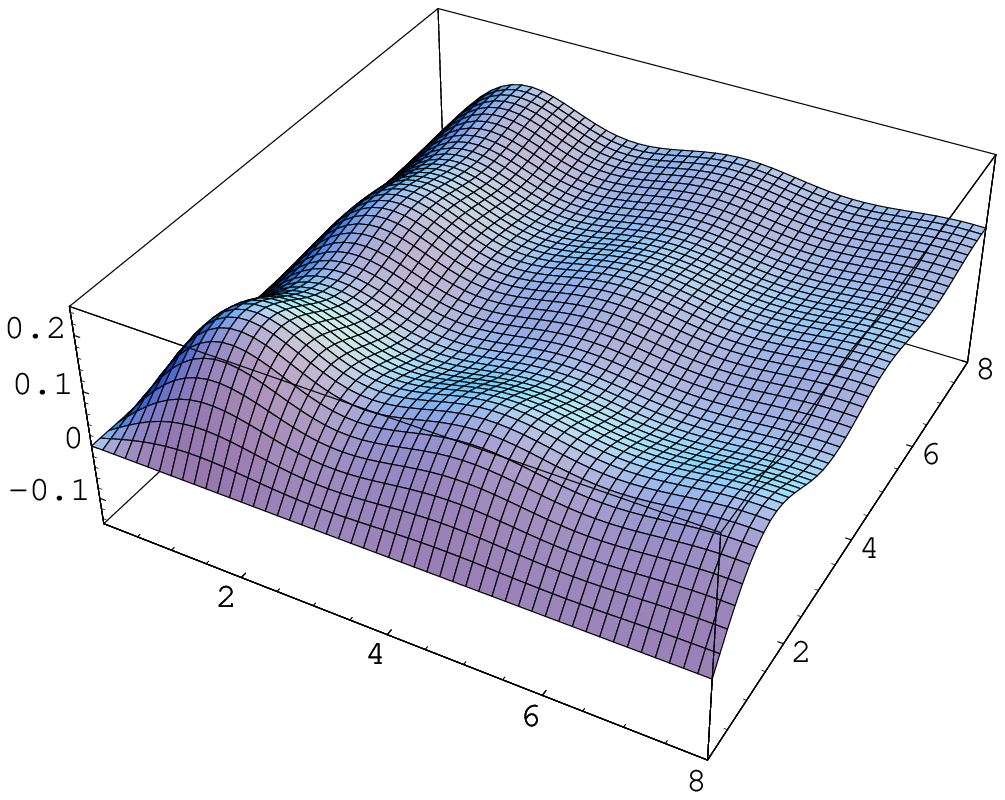,width=15pc}
\epsfig{figure=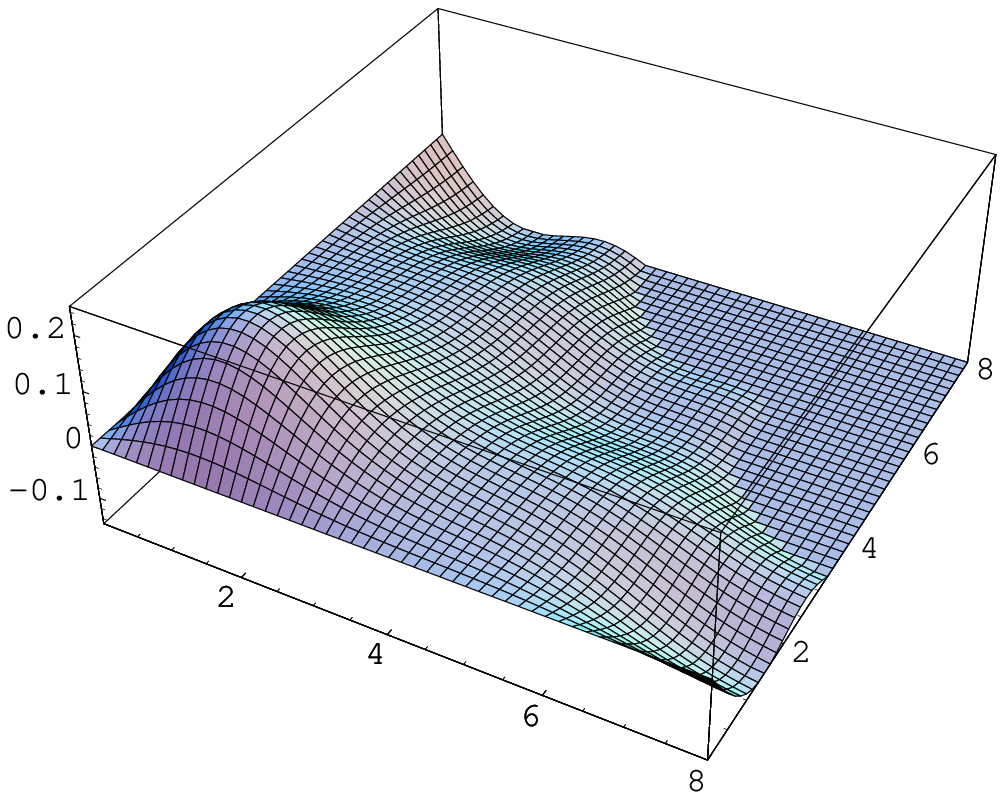,width=15pc}
\put(-360,140){$\rho_{1,1}(s,t)$}
\put(-180,140){$p_{1,1}(s,t)$ eq.(16)}
}
\centerline{\epsfig{figure=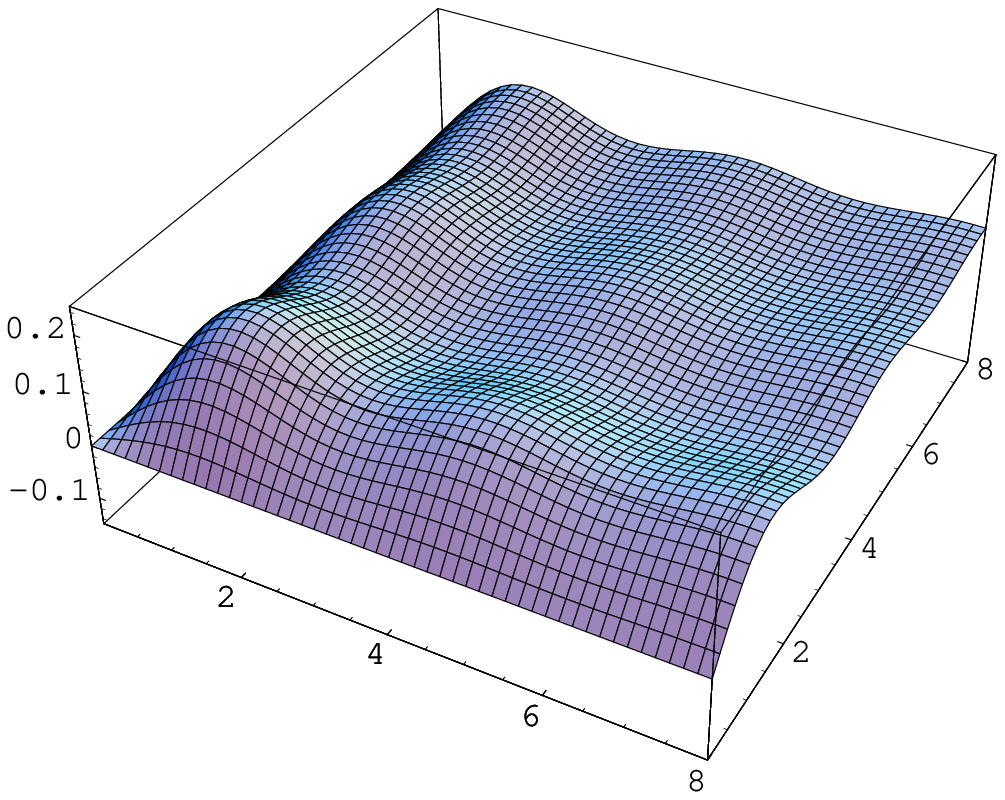,width=15pc}
\epsfig{figure=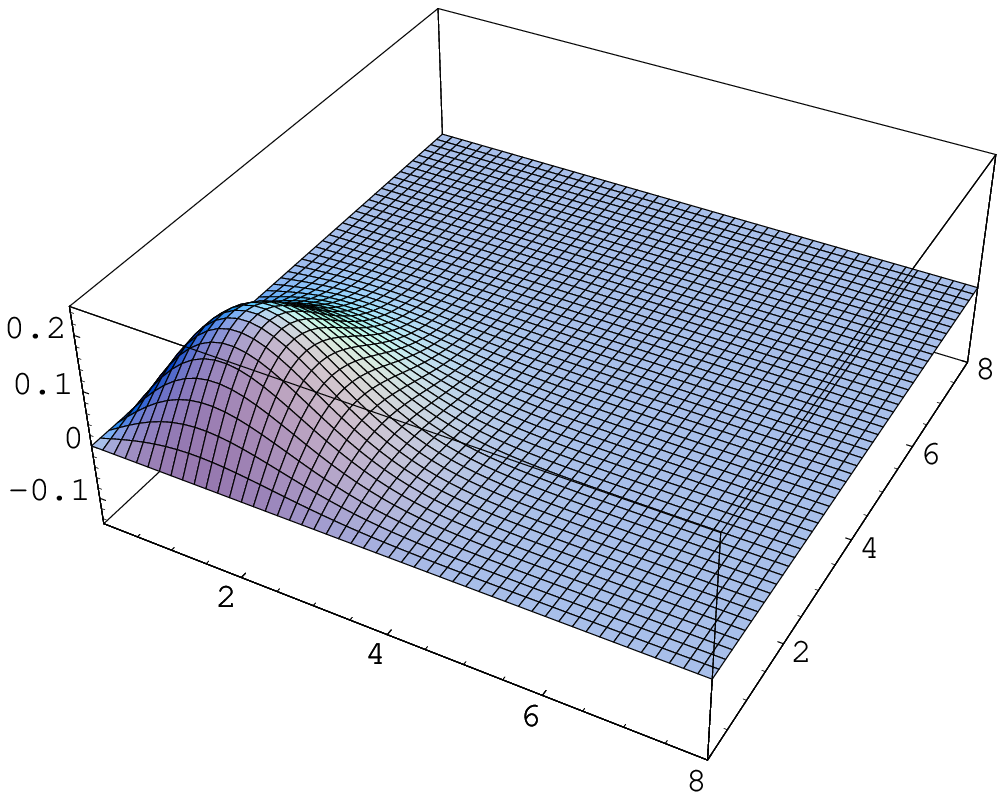,width=15pc}
\put(-360,130){$\rho_1(s)\rho_1(t)$}
\put(-170,130){$p_1(s)p_1(t)$}
}
\caption{Factorisation of the quenched
density $\rho_{1,1}(s,t)$: eq. (\protect{\ref{r11}}) 
at $\nu=0$ and $\alpha=4.318$ 
(top left) vs. the factorised eq. (\protect{\ref{factorR}}), using  
eq. (\protect{\ref{rhoQ}}) (bottom left). The corresponding approximate 
individual eigenvalue distribution $p_{1,1}(s,t)$ eq. (\protect{\ref{p00ex}}) 
for the same values (top right) vs. the factorised exact result
eq. (\protect{\ref{pklfac}}) using eq.  (\protect{\ref{p1}}) (bottom right). 
}
\label{R11factor}
\end{figure} 
\bea
\lim_{\alpha\gg1}E_{k,l}(s,t) 
&=& \left(\sum_{i=0}^{N-k}\frac{(-)^i}{i!}\int_0^s  dx_1\ldots dx_{k+i}\
\rho_{k+i,0}^{(0)}(x_1,\ldots,x_{k+i})\right)\nn\\
&&\times
\left(
\sum_{j=0}^{N-l}\frac{(-)^j}{j!} 
\int_0^t  dy_1\ldots dy_{l+j}\ 
\rho_{0,l+j}^{(0)}(y_1,\ldots,y_{l+j})\right)\nn\\
&=& E_{k}(s) E_{l}(t) 
\label{EklRfac}
\eea
Differentiating twice as in eq. (\ref{Eklpkl}) we get
\be
 \frac{\partial^2E_{k,l}(s,t)}{\partial s\partial_t} 
=  \frac{\partial E_{k}(s)}{\partial s} 
 \frac{\partial E_{l}(t)}{\partial_t} =
k! (p_k(s)-p_{k+1}(s))\ l!  (p_l(t)-p_{l+1}(t)),
\ee
and thus from comparing to eq. (\ref{Eklpkl})
\be
p_{k,l}(s,t)=p_k(s)p_l(t)\ .
\label{pklfac}
\ee
The 1RMT quantities on the right hand side are now explicitly known, without
approximations. The comparison in 3D is given in fig. \ref{R11factor} right. 
In order to check the convergence of eq. (\ref{p00ex}) we can cut the 3D plot
and compare to the exact factorised result, as shown in fig. 
\ref{R1p1exactfactor}.
\begin{figure}[-h]
\includegraphics[width=.5\textwidth]{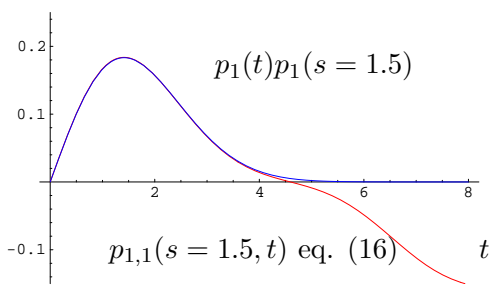}
\put(-100,80){$p_{1}(t)p_1(s=1.5)$}
\put(-140,10){$p_{1,1}(s=1.5,t)$ eq. (16)}
\put(0,10){$t$}
\caption{2D cut of fig.  \protect{\ref{R11factor}} top right at $s=1.5$ (red) 
vs. the exact factorised distribution eq. (\protect{\ref{pklfac}}) 
(blue). 
}
\label{R1p1exactfactor}
\end{figure}

%%%%%%%%%%%%%%%%%%%%%%%%%%%%%%%%%%%%%%%%%%%%%%%%%%%%%%%%%%%%%%%%%%%%%%%%%%%

\subsection{Factorisation of the weight function}\label{wfac}

In this section we show that the bi-orthogonal polynomials for the
non-Gaussian  
weight eq. (\ref{weight}) can be constructed in terms of orthogonal
polynomials with Gaussian weight. Our discussion follows closely appendix B of
\cite{ADOS}. Suppose we have two sets of ordinary orthogonal polynomials
\bea
\int dx ~ w_1(x) ~ P_k(x^2) ~ P_l(x^2) &=&  f_k \delta_{k,l} \nn \\
\int dy ~ w_2(y) ~ Q_k(y^2) ~ Q_l(y^2) &=&  g_k \delta_{k,l} ~,
\eea
with weights $w_{1,2}$ and norms $f_k$ and $g_k$ respectively. Then it follows
that these polynomials are bi-orthogonal with respect to the weight
\bea
w(x,y) &\equiv& w_1(x) ~ w_2(y) ~
 \sum_{k=0}^\infty ~ \frac{h_k}{f_k g_k} ~ P_k(x~^2) ~ Q_k(y^2) ~.
\label{wexp-w}
\eea
In our quenched case we can simply choose the ordinary Laguerre weight and its
polynomials, 
\bea
w_1(x) &=& w_2(x)\ =\ x^{2\nu+1} e^{-\frac{N}{1+\mu^2}x^2}\nn\\
P_k(x^2) &=& Q_k(x^2)\ \sim\ 
%k! \frac{(1+\mu^2)^k}{(-N)^k} 
L_k^{(\nu)} \left(\frac{N}{1+\mu^2}x^2\right).
\label{wexp-wop}
\eea
The identity that allows to link the Laguerre weight and 
polynomials to the non-Gaussian
weight eq. (\ref{weight}) is given by
\bea
w(x,y) &=& (xy)^{\nu+1} ~ e^{-\frac{N(1+\mu^2)}{4\mu^2}(x^2+ y^2)} ~
  I_{\nu}\left(\frac{1-\mu^2}{2\mu^2}N x y\right) 
\label{B7}\\
 &=& 4\mu^2\frac{(N(1-\mu^2))^\nu}{(1+\mu^2)^{2\nu+1}} 
(x y)^{2\nu+1} e^{-\frac{N}{1+\mu^2} (x^2 +y^2)}\nn\\
&&\times \sum_{n=0}^{\infty} \frac{n!(1-\tau)^n}{(n+\nu)!} 
 L_n^{(\nu)} \left(\frac{N}{1+\mu^2} x^2\right) 
L_n^{(\nu)} \left(\frac{N}{1+\mu^2}y^2\right), 
\nn
\eea
satisfying eq. (\ref{wexp-w}).
The Bessel transforms eq. (\ref{chidef}) then simply result into the wave
functions 
\bea
\label{wexp-bt}
\chi_k(y) &=& \int dx ~ w(x,y) ~ P_k(x^2) 
\ \sim\ 
%\frac{2\,k!\mu^2(1+\mu^2)^{k-1}}{-(-N)^{k+1}}
%\left(\frac{1-\mu^1}{1+\mu^2}\right)^{2k+\nu}\!\!\!
y^{2\nu+1}e^{-N\frac{N}{1+\mu^2}y^2}
L_k^{(\nu)} \left(\frac{N}{1+\mu^2}y^2\right)\!.\ 
\eea
All the kernels can now be easily written in terms of Laguerre polynomials and
their norms, and we refer to \cite{ADOS} for details. 
Finally let us reconsider the upper left block in eq. (\ref{allRnk}). Due to
the above identity eq. (\ref{B7}) we obtain
\beq
w(x,y) - M_N(x,y) ~=~  w_1(x) ~ w_2(y) ~
 \sum_{k=N}^\infty ~ \frac{h_k}{f_k g_k} \ P_k(x^{2})\   Q_k(y^2) .
\label{w-M}
\eeq
Thus naively taking the limit $N\to\infty$
we would expect the right hand side to
vanish. However, in the limit eq. (\ref{limN}) this is not the case, and we
instead obtain $M_N(x,y)\to {\cal I}^-(\xi,\zeta)$. 
Only in the limit of large $\alpha\gg1$ limit  
the integral in ${\cal I}^-$ extends to $\infty$ in the new variables
$\xi/\sqrt{\alpha}$, making it converge to the weight (see eq. (\ref{Mtow})). 
This explains the factorisation in this limit, illustrating the subtlety of the
large-$N$ limit (that is distinguished into weak and strong non-Hermiticity for
real $\mu$).\\

\indent

{\bf Acknowledgements:} 

\noindent
I would like to thank the organisers for their generous hospitality during 
this very stimulating workshop. It is a pleasure to thank F. Basile, 
P. Damgaard, J. Osborn and K. Splittorff with whom the results have been
obtained that are covered in this talk. 
This work was supported by 
EPSRC grant EP/D031613/1 and 
EU network ENRAGE MRTN-CT-2004-005616.


\newpage 

\begin{thebibliography}{99}

\bibitem{Amu07}
  G.~Akemann,
  %``Matrix models and QCD with chemical potential,''
  Int.\ J.\ Mod.\ Phys.\  A {\bf 22} (2007) 1077.
%  [arXiv:hep-th/0701175].

\bibitem{Jackra}
K.~Splittorff and J.~J.~M.~Verbaarschot, arXiv:0710.0704 [hep-th].

\bibitem{KimJac03}
K.~Splittorff and J.~J.~M.~Verbaarschot,
  %``Factorization of correlation functions and the replica limit of the  Toda
  %lattice equation,''
  Nucl.\ Phys.\ B {\bf 683} 467 (2004)
  [hep-th/0310271].
  %%CITATION = HEP-TH 0310271;%%

\bibitem{James}
J.~C.~Osborn,
  %``Universal results from an alternate random matrix model for QCD with a
  %baryon chemical potential,''
  Phys.\ Rev.\ Lett.\  {\bf 93}, 222001 (2004)
  [hep-th/0403131].
  %%CITATION = HEP-TH 0403131;%%

\bibitem{AOSV}
G.~Akemann, J.~C.~Osborn, K.~Splittorff and J.~J.~M.~Verbaarschot,
%``Unquenched QCD Dirac operator spectra at nonzero baryon chemical
%potential,''
Nucl.\ Phys.\ B {\bf 712} (2005) 287
[hep-th/0411030];
%%CITATION = HEP-TH 0411030;%%

\bibitem{ABSW}
G.~Akemann, J.~Bloch L.~Shifrin and T.~Wettig, PoS(Lattice2007)244.

\bibitem{KimJac07} K.~Splittorff and J.~J.~M.~Verbaarschot, Phys. Rev. Lett. 
{\bf 98} (2007) 031601 [hep-lat/0609076]; Phys. Rev. {\bf D75} (2007) 116003
[hep-lat/0702011].

\bibitem{AW} G. Akemann and T. Wettig, 
Phys. Rev. Lett. {\bf 92} (2004) 102002
\{Erratum-ibid.
{\bf 96} (2006) 029902\}
  [hep-lat/0308003];
J.~C.~Osborn and T.~Wettig,
  %``Dirac eigenvalue correlations in quenched QCD at finite density,''
  PoS (LAT2005) 200
  [hep-lat/0510115];
%\bibitem{Bloch:2006cd}
  J.~Bloch and T.~Wettig,
  %``Overlap Dirac operator at nonzero chemical potential and random matrix
  %theory,''
  Phys.\ Rev.\ Lett.\  {\bf 97} (2006) 012003 [hep-lat/0604020].
%  [arXiv:hep-lat/0604020].

\bibitem{KimBen} K. Splittorff and B. Svetitsky, Phys. Rev. {\bf D75}
(2007) 114504 [hep-lat/0703004].

\bibitem{BA}
F.~Basile, G.~Akemann, archive/0710.0376 [hep-th].

\bibitem{GL}
J. Gasser and H.~Leutwyler,
  Phys.\ Lett.\ {\bf B188} (1987) 477.

\bibitem{DomJac} 
D. Toublan and J.J.M. Verbaarschot,
Nucl. Phys. {\bf B603} (2001) 343 %-368
[hep-th/0012144].

\bibitem{DHSST}
P.~H.~Damgaard, U.~M.~Heller, K.~Splittorff and B.~Svetitsky,
  %``A new method for determining F(pi) on the lattice,''
  Phys.\ Rev.\ D {\bf 72} (2005) 091501
  [hep-lat/0508029];
  %%CITATION = HEP-LAT 0508029;%%
%
%\bibitem{DHSST}
P.~H.~Damgaard, U.~M.~Heller, K.~Splittorff, B.~Svetitsky and D.~Toublan,
  %``Extracting F(pi) from small lattices: Unquenched results,''
  Phys.\ Rev.\ D {\bf 73} (2006) 074023
  [hep-lat/0602030];
  %%CITATION = HEP-LAT 0602030;%%
  %``Microscopic eigenvalue correlations in QCD with imaginary isospin chemical
  %potential,''
  Phys.\ Rev.\ D {\bf 73} (2006) 105016
  [hep-th/0604054].
  %%CITATION = HEP-TH 0604054;%%

\bibitem{ADOS}
G.~Akemann, P.~H.~Damgaard, J.~C.~Osborn and K.~Splittorff,
  %``A new chiral two-matrix theory for Dirac spectra with imaginary chemical
  %potential,''
Nucl. Phys. {\bf B766} (2007) 34 
[hep-th/0609059].
  %%CITATION = HEP-TH/0609059;%%

\bibitem{DeG} T. DeGrand and S. Schaefer, PoS(Lattice2007)069;
archive/0708.1731v1 [hep-lat].

\bibitem{ADlat07}
G.~Akemann and P.~H.~Damgaard, PoS(Lattice2007)166, 
arXiv:0709.0484v1 [hep-lat].

\bibitem{DNW}
  S.~M.~Nishigaki, P.~H.~Damgaard and T.~Wettig,
  %``Smallest Dirac eigenvalue distribution from random matrix theory,''
  Phys.\ Rev. {\bf D58} (1998) 087704
  [hep-th/9803007];

\bibitem{ADp}
  G.~Akemann and P.~H.~Damgaard,
  %``Distributions of Dirac operator eigenvalues,''
  Phys.\ Lett. {\bf B583} (2004) 199
  [hep-th/0311171].
  %%CITATION = HEP-TH 0311171;%%


\end{thebibliography}
\end{document}